\documentclass[showpacs,aps,twocolumn,prb]{revtex4-1}
\usepackage{amsfonts}
\usepackage{graphicx}
\usepackage{amsmath}
\usepackage{amssymb}

\begin{document}

\title{Hexagonal warping on spin texture, Hall conductivity and circular
dichroism of Topological Insulator}
\author{Zhou Li$^1$}
\email{lizhou@mcmaster.ca}
\author{J. P. Carbotte$^{1,2}$}
\email{carbotte@mcmaster.ca}
\affiliation{$^1$ Department of
Physics, McMaster University, Hamilton, Ontario,
Canada,L8S 4M1 \\
$^2$ Canadian Institute for Advanced Research, Toronto, Ontario,
Canada M5G 1Z8}

\begin{abstract}
The topological protected electronic states on the surface of a
topological insulator can progressively change their Fermi
cross-section from circular to a snowflake shape as the chemical
potential is increased above the Dirac point because of an hexagonal
warping term in the Hamiltonian. Another effect of warping is to
change the spin texture which exists when a finite gap is included
by magnetic doping, although the in-plane spin component remains
locked perpendicular to momentum. It also changes the orbital
magnetic moment, the matrix element for optical absorption and the
circular dichroism. We find that the Fermi surface average of
z-component of spin is closely related to the value of the Berry
phase. This holds even when the Hamiltonian includes a subdominant
non-relativistic quadratic in momentum term (which provides
particle-hole asymmetry) in addition to the dominant relativistic
Dirac term. There is also a qualitative correlation between
$\left\langle S_z \right\rangle$ and the dichroism. For the case
when the chemical potential falls inside the gap between valence and
conduction band, the Hall conductivity remains quantized and
unaffected in value by the hexagonal warping term.
\end{abstract}

\pacs{75.70.Tj,78.67.-n,73.43.Cd }
\date{\today }
\maketitle

\section{Introduction}
Helical Dirac fermions exist at the surface of a topological insulator.
There is an odd number of Dirac points in the surface state two dimensional
honeycomb lattice Brillouin zone which are topologically protected. \cite%
{Hasan,Qi1,Moore,Hsieh1,Chen} Spin and angular resolved
photoemission (ARPES) reveals spin-momentum ($\mathbf{k}$) locking
\cite{Hsieh2} with in plane spin component perpendicular to
$\mathbf{k}$. Another important feature of the helical Dirac
fermions observed in topological insulators is that the fermi
contours are circular for small values of the chemical potential
($\mu $) in the conduction band and acquire a snowflake \cite{Chen}
shape as $\mu $ increases. This has been assigned by Fu \cite{Fu} to
a hexagonal warping term in the Hamiltonian of such charge carriers.
This term has a strong signature in the optical conductivity.
\cite{Li1} The usual flat background
\cite{Carbotte1,Carbotte2,Carbotte3,Li08,Orlita} associated with the
interband transitions in graphene is predicted to show instead a
large near linear increase with photon energy above the interband
threshold. It is also possible to introduce a gap in the helical
electrons by magnetic doping \cite{Chen1} of the surface states to
break time reversal symmetry and thus produce massive Dirac fermions
in $Bi_{2}Se_{3}$. \cite{Chen1} Finally we note that the Dirac
spectrum at the surface of a TI usually displays considerable
particle-hole asymmetry which can be modeled with a small
sub-dominant Schr\"{o}dinger quadratic in momentum kinetic energy
piece to the Hamiltonian which is in addition to the dominant Dirac
piece. The asymmetry provides an hourglass or goblet shape to the
valence band dispersion curves. \cite{Hsieh2} While perhaps small,
the Schr\"{o}dinger piece has been shown to provide important
modifications \cite{Li2} in the magneto optics of such systems as
compared with what is found in graphene \cite{Carbotte4,Pound} or
the related single layer silicene. \cite{Tabert} In this paper we
will be primarily interested in the effect of hexagonal warping on
the Berry curvature, orbital magnetic moment, Berry phase, in and
out of plane spin texture, matrix elements for optical absorption
and on the circular dichroism, we will also consider the effect of
including a gap as a primary element and of a subdominant non
relativistic Schr\"{o}dinger contribution to the Hamiltonian.

The paper is structured as follows. In section II we introduce the
Hamiltonian for the helical Dirac electrons on the surface of a
topological insulator. It includes a Dirac piece which involves real
spin, a hexagonal warping term and a Schr\"{o}dinger contribution to
the kinetic energy quadratic in momentum with effective mass $m$. We
also include the possibility of a gap opening. Berry curvature and
orbital magnetic moment are described. Section III considers the
Berry phase of such a system and in particular how it is changed by
the warping and quadratic in momentum Schr\"{o}dinger term. A
discussion of spin texture is given in section IV. A strong
correlation between the Fermi surface average of the z-component of
spin and the Berry phase of the corresponding orbit is established.
Section V describes optical absorption of circular polarized light
and dichroism. The Hall conductivity is addressed, and we relate the
dichroism to the optical matrix elements for circularly polarized
light. Because of the anisotropy introduced by the warping it is
essential to average the optical matrix elements associated with the
Hall and longitudinal conductivity separately, before taking their
ratio to obtain the Hall angle. A summary and conclusions are found
in section VI.
\section{Hamiltonian and orbital magnetic moment}

\begin{figure}[tp]
\begin{center}
\includegraphics[height=3.9in,width=2.5in]{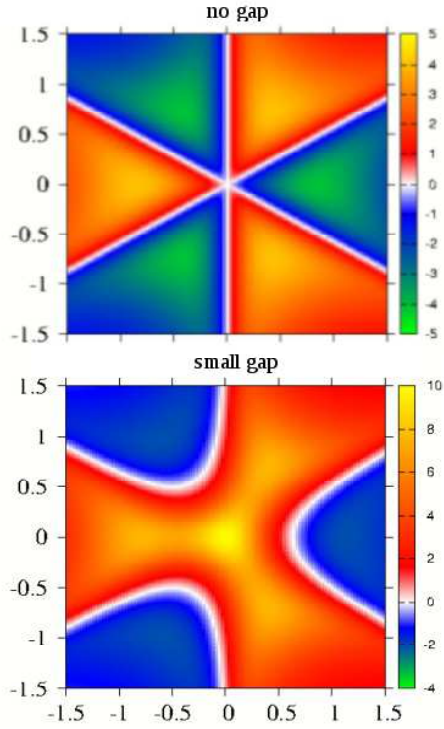}
\end{center}
\caption{(Color online) Color plot of the orbital magnetic moment
$m_{z}$ as a function of $k_x$ and $k_y$ in the surface state
Brillouin zone with momentum in units of $nm^{-1}$. The top frame is
for $\Delta=0$ and the
bottom for $\Delta=0.1eV$. In both frames the hexagonal warping $\protect%
\lambda=0.2$ in unit of $eV\cdot nm^{3}$.} \label{fig1}
\end{figure}

\begin{figure}[tp]
\begin{center}
\includegraphics[height=2.8in,width=2.8in,angle=-90]{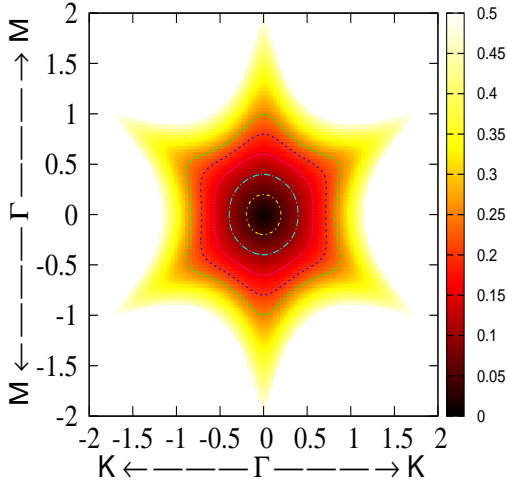}
\end{center}
\caption{(Color online) Constant energy contours C for various
values of the chemical potential $\protect\mu$ as a function of
momentum $k_x$, $k_y$ in the surface states Brillouin zone with
momentum in units of $nm^{-1}$.} \label{fig2}
\end{figure}
The Hamiltonian used by Fu \cite{Fu} to describe the surface states
band structure near the $\Gamma $ point in the surface Brillouin
zone of a topological insulator is
\begin{equation}
H_{0}=\hbar v_{k}(k_{x}\sigma _{y}-k_{y}\sigma _{x})+\frac{\lambda }{2}%
(k_{+}^{3}+k_{-}^{3})\sigma _{z}+\Delta \sigma _{z}+E_{0}(\mathbf{k})
\label{Hamiltonian}
\end{equation}%
where $E_{0}(\mathbf{k})=\hbar ^{2}k^{2}/(2m)$ is a quadratic term which
gives the Dirac fermionic dispersion curves an hour glass shape and provides
particle-hole asymmetry. The Dirac fermion velocity to second order is $%
v_{k}=v_{F}(1+\alpha k^{2})$ with $v_{F}$ the usual Fermi velocity measured
to be $2.55eV\cdot \mathring{A}$ and $\alpha $ is a constant which is fit
along with $m$ to the measured band structure in reference (\onlinecite{Fu}%
). The hexagonal warping parameter $\lambda =250eV\cdot \mathring{A}^{3}$.
The $\sigma _{x}$, $\sigma _{y}$, $\sigma _{z}$ are the Pauli matrices here
referring to spin, while in graphene these would relate instead to
pseudospin. Finally $k_{\pm }=k_{x}\pm ik_{y}$ with $k_{x}$, $k_{y}$
momentum along $x$ and $y$ axis respectively. The energy spectrum associated
with the Hamiltonian [Eq.~(\ref{Hamiltonian})] is
\begin{equation}
\varepsilon _{\pm }(\mathbf{k})=E_{0}(\mathbf{k})\pm \sqrt{\hbar
^{2}v_{k}^{2}k^{2}+(\lambda k^{3}\cos (3\theta )+\Delta )^{2}}  \label{disp}
\end{equation}%
where $\theta $ is the polar angle defining the direction of $\mathbf{k}$ in
the two dimensional surface state Brillouin zone. The wave function $u(%
\mathbf{k},s)$ defined by the equation $H_{0}u(\mathbf{k},\pm )=\varepsilon
_{\pm }(\mathbf{k})u(\mathbf{k},\pm ),$ is given by
\begin{equation}
u(\mathbf{k},s)=\frac{\hbar v_{k}k\left( 1,\frac{\Delta (k,\theta )-s\sqrt{%
\hbar ^{2}v_{k}^{2}k^{2}+(\Delta (k,\theta ))^{2}}}{\hbar v_{k}}\frac{%
-ik_{x}+k_{y}}{k^{2}}\right) ^{T}}{\sqrt{\hbar ^{2}v_{k}^{2}k^{2}+(\Delta
(k,\theta )-s\sqrt{\hbar ^{2}v_{k}^{2}k^{2}+(\Delta (k,\theta ))^{2}})^{2}}}
\label{wave}
\end{equation}%
Here $s=+/-$ gives conduction and valence band respectively and $\Delta
(k,\theta )=\Delta +\lambda k^{3}\cos (3\theta )\equiv \Delta +\lambda
(k_{x}^{3}-3k_{x}k_{y}^{2})$. Note that the quadratic in momentum Schr\"{o}%
dinger term $E_{0}(\mathbf{k})$ does not appear in the wave function.
Introducing a unit vector $\hat{z}$ perpendicular to the surface plane, the
Berry curvature associated with the wave function (for simplicity we set $%
v_{k}=v$) is given by
\begin{eqnarray}
&&\Omega _{c}(\mathbf{k})=\hat{z}\cdot \nabla _{\mathbf{k}}\times
\left\langle u(\mathbf{k},+)|i\nabla _{\mathbf{k}}|u(\mathbf{k}%
,+)\right\rangle  \notag \\
&=&-\frac{\hbar ^{2}v^{2}(\Delta -2\lambda k_{x}^{3}+6\lambda k_{x}k_{y}^{2})%
}{2[(\Delta +\lambda (k_{x}^{3}-3k_{x}k_{y}^{2}))^{2}+\hbar
^{2}v^{2}k^{2}]^{3/2}}  \label{curve}
\end{eqnarray}%
for the conduction band and for the valence band $\Omega _{v}(\mathbf{k}%
)=-\Omega _{c}(\mathbf{k})$.

Closely related to the Berry curvature is the orbital magnetic moment $m_{z}(%
\mathbf{k},s)$ which for the conduction band $s=+$ is given by \cite{Xiao1}
\begin{eqnarray}
&&m_{z}(\mathbf{k,+})  \notag \\
&=&-i(e/2\hbar )\{\hat{z}\cdot \left\langle \nabla _{\mathbf{k}}u(\mathbf{k}%
,+)|\times \lbrack H_{0}-\varepsilon _{+}(k)]|\nabla _{\mathbf{k}}u(\mathbf{k%
},+)\right\rangle \}  \notag \\
&=&-(e/2\hbar )\frac{\hbar ^{2}v^{2}(\Delta -2\lambda k_{x}^{3}+6\lambda
k_{x}k_{y}^{2})}{[(\Delta +\lambda (k_{x}^{3}-3k_{x}k_{y}^{2}))^{2}+\hbar
^{2}v^{2}k^{2}]}  \label{mz}
\end{eqnarray}%
and for the valence band $m_{z}(\mathbf{k},-)=m_{z}(\mathbf{k,+})$. Except
for numerical factors this expression differs from the Eq.~(\ref{curve}) for
the Berry curvature only in its denominator which appears to power one
rather than $3/2$. When we do not include a warping term in the Hamiltonian
the Berry curvature as well as the orbital magnetic moment is proportional
to the gap and would vanish for $\Delta =0$. This no longer is the case if
warping is included. Even with $\Delta =0$ there is an orbital magnetic
moment as we show in Fig.~\ref{fig1} which is a color plot of the magnitude
of $m_{z}$ as a function of momentum ($k_{x}$, $k_{y}$) in the surface state
Brillouin zone with momentum in units of $\pi $. We see that $%
m_{z}(k_{x},k_{y})$ is finite in most of the $k_{x}$-$k_{y}$ plane with zero
along the lines of $\theta =\pm \pi /6$ and $\theta =\pi /2$. Thus $%
m_{z}(k_{x},k_{y})$ changes sign 6 times as $\theta $ ranges from $0$ to $%
2\pi $. This is in sharp contrast to the case when the hexagonal warping
term is zero but the gap is finite. In this case $m_{z}(k,+)$ reduces to
\begin{equation}
-\frac{e}{2\hbar }\frac{\hbar ^{2}v^{2}\Delta }{\Delta ^{2}+\hbar
^{2}v^{2}k^{2}}
\end{equation}
which is isotropic in the $(k_{x},k_{y})$ plane and is peaked at $k=0$. In
fact in the limit $\Delta \rightarrow 0$ we get $-\frac{e\pi }{2\hbar }%
\delta (k^{2})$. This maximum around the origin remains even when warping is
included as we show in the color plot in the lower frame of Fig.~\ref{fig1}
where a small gap $\Delta =0.1eV$ is included alongside a finite $\lambda $.
The same value of $\lambda =0.2eV\cdot nm^{3}$ was used in both top and
bottom frame. We note that now the contours of zero orbital magnetic moment
are no longer straight lines and that we no longer have perfect symmetry
between regions of positive and negative magnetic moments. The zero along $%
\theta=0,\pm2\pi/3$ are at a finite value of the absolute momentum $%
k=(\Delta/(2\lambda))^{1/3}$ which represents the minimum value of momentum
for which the magnetic moment can vanish whatever the direction of $\mathbf{k%
}$.

\section{Berry phase}

\begin{figure}[tp]
\begin{center}
\includegraphics[height=3.0in,width=2.8in]{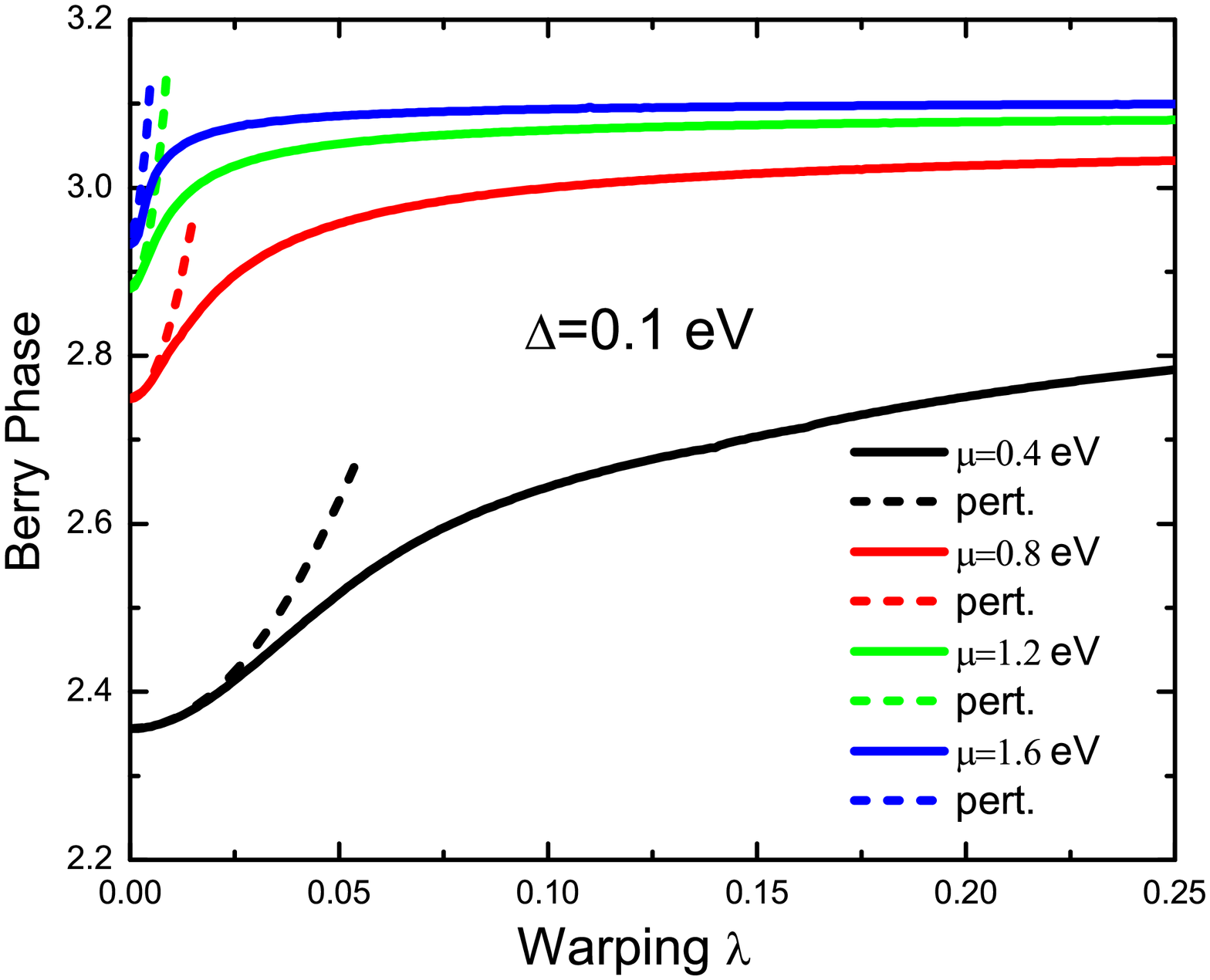}
\end{center}
\caption{(Color online) Effect of warping on Berry phase. The Schr\"{o}%
dinger term $E_0(k)=0$, the gap $\Delta=0.1eV$. Several values of chemical
potential are considered and color coded. In all cases the dashed lines are
our simplified results of Eq.~(\protect\ref{Berry}) for the small $\protect%
\lambda$ limit. The horizontal axis is $\protect\lambda$ (warping) in unit
of $eV\cdot nm^{3}$. The solid lines are from Eq.~(\protect\ref{Berry1})
evaluated numerically.}
\label{fig3}
\end{figure}

Based on the Berry curvature we can also calculate the Berry phase for a
closed contour C, which is defined as%
\begin{equation}
\Gamma _{s}(C)=\oint_{C}d\mathbf{k\cdot }\left\langle u(\mathbf{k}%
,s)|i\nabla _{\mathbf{k}}|u(\mathbf{k},s)\right\rangle\label{Berryf}
\end{equation}%
By Stokes' theorem we know that%
\begin{equation}
\oint_{C}d\mathbf{k\cdot A=}\iint_{S}dk_{x}dk_{y}(\nabla \times \mathbf{%
A\cdot (-}\hat{z}))
\end{equation}%
We assume that the contour C is clockwise and the area enclosed by it is
pointing in the negative $\hat{z}$ direction, and the chemical potential is
positive for definiteness. Thus
\begin{eqnarray}
&&\Gamma _{+}(C)=\iint_{S}dk_{x}dk_{y}\Theta (\mu -\varepsilon _{+}(\mathbf{k%
}))  \notag \\
&&\times \frac{\hbar ^{2}v^{2}(\Delta -2\lambda k_{x}^{3}+6\lambda
k_{x}k_{y}^{2})}{2[(\Delta +\lambda (k_{x}^{3}-3k_{x}k_{y}^{2}))^{2}+\hbar
^{2}v^{2}k^{2}]^{3/2}}  \label{phase}
\end{eqnarray}%
where $\mu $ is the energy for the contour C. To see the effect of hexagonal
warping on the Berry phase we start with no Schr\"{o}dinger term $E_{0}(k)=0$
in Eq.~(\ref{Hamiltonian}). We get
\begin{eqnarray}
&&\Gamma _{+}(C)=\int_{0}^{2\pi }d\theta \int_{0}^{k_{F}(\theta )}kdk  \notag
\\
&&\times \frac{\hbar ^{2}v^{2}(\Delta -2\lambda k^{3}\cos (3\theta ))}{%
2[(\Delta +\lambda k^{3}\cos (3\theta ))^{2}+\hbar ^{2}v^{2}k^{2}]^{3/2}}
\end{eqnarray}%
The integration over the magnitude of $k$ can be performed using the
integral
\begin{equation}
\int \frac{(B-2Ak^{3})kdk}{[k^{2}+(B+Ak^{3})^{2}]^{3/2}}=\frac{-(B+Ak^{3})}{%
\sqrt{k^{2}+(B+Ak^{3})^{2}}}
\end{equation}%
and we get
\begin{eqnarray}
&&\Gamma _{+}(C)  \notag \\
&=&\int_{0}^{2\pi }d\theta \frac{-(\Delta +\lambda k^{3}\cos (3\theta ))}{%
2[(\Delta +\lambda k^{3}\cos (3\theta ))^{2}+\hbar ^{2}v^{2}k^{2}]^{1/2}}%
|_{0}^{k_{F}(\theta )}  \notag \\
&=&\frac{1}{2}\int_{0}^{2\pi }d\theta \{sign(\Delta )-\frac{(\Delta +\lambda
k_{F}^{3}(\theta )\cos (3\theta ))}{\mu }\}  \label{Berry1}
\end{eqnarray}%
where $k_{F}(\theta )$ is determined by
\begin{equation*}
\sqrt{(\Delta +\lambda k_{F}^{3}(\theta )\cos (3\theta ))^{2}+\hbar
^{2}v^{2}k_{F}^{2}(\theta )}=\mu .
\end{equation*}
Several limits of this expression should be emphasized. First in the limit
of $\mu \rightarrow \infty $ we get a Berry phase of $\pi sign(\Delta )$ and
this is the same value as when $\lambda =0$ i.e. no warping is included.
While warping does not change the Berry phase when the contour has $\mu
\rightarrow \infty $ it does change it for any finite $\mu $. We can get an
analytic expression for this change in the limit of $\lambda \rightarrow 0$
retaining lowest order.

To evaluate the Berry phase Eq.~(\ref{Berry1}) we need to know the fermi
momentum $k_{F}$ as a function of angle $\theta $. We show such contours in
Fig.~\ref{fig2} for various values of $\mu $. At small $\mu $ the contour is
nearly circular and distorts into a snowflake shape as $\mu $ increases. We
can solve for $k_{F}(\theta )$ v.s. angle $\theta $. Keeping only lowest
significant order in $\lambda $ for small $\lambda $, the equation to be
solved is
\begin{equation}
(\Delta +\lambda k_{F}^{3}(\theta )\cos (3\theta ))^{2}+\hbar
^{2}v^{2}k_{F}^{2}(\theta )=\mu ^{2}  \label{chemical}
\end{equation}%
which is a cubic equation for $k_{F}(\theta )$. For $\lambda =0$ (no
warping) $k_{F}$ is isotropic and equals $\sqrt{\mu ^{2}-\Delta ^{2}}/(\hbar
v)$. We can use this in Eq.~(\ref{chemical}) to get a first order correction
to $k_{F}(\theta )$
\begin{equation}
\hbar ^{2}v^{2}\frac{k_{F}^{2}(\theta )}{\mu ^{2}}\cong 1-\frac{\Delta ^{2}}{%
\mu ^{2}}-2\frac{\Delta }{\mu }\frac{\lambda }{\mu }(\frac{\mu ^{2}-\Delta
^{2}}{\hbar ^{2}v^{2}})^{3/2}\cos (3\theta )
\end{equation}%
from which we obtain
\begin{eqnarray}
&&k_{F}^{3}(\theta )\cong (\frac{\mu }{\hbar v})^{3}(\frac{\mu ^{2}-\Delta
^{2}}{\mu ^{2}})^{3/2}  \notag \\
&&\times \lbrack 1-3\frac{\lambda \mu \Delta }{(\hbar v)^{3}}\sqrt{\frac{\mu
^{2}-\Delta ^{2}}{\mu ^{2}}}\cos (3\theta )]
\end{eqnarray}%
This gives a Berry phase
\begin{eqnarray}
&&\Gamma _{+}(C)=\frac{1}{2}\int_{0}^{2\pi }d\theta \{sign(\Delta )-\frac{%
\Delta }{\mu }-\frac{\lambda }{\mu }\cos (3\theta )(\frac{\mu }{\hbar v})^{3}
\notag \\
&&\times (\frac{\mu ^{2}-\Delta ^{2}}{\mu ^{2}})^{3/2}[1-3\frac{\lambda \mu
\Delta }{(\hbar v)^{3}}\sqrt{\frac{\mu ^{2}-\Delta ^{2}}{\mu ^{2}}}\cos
(3\theta )]\}
\end{eqnarray}%
The one in the last bracket will give zero after integration over angle $%
\theta $ because this term is linear in $\cos (3\theta )$ which averages to
zero. The second term however involves $\int_{0}^{2\pi }d\theta \cos
^{2}(3\theta )$ which is non-zero and equal to $\pi $. Therefore
\begin{equation}
\Gamma _{+}(C)=\pi \lbrack sign(\Delta )-\frac{\Delta }{\mu }(1-\frac{3}{2}%
\frac{\lambda ^{2}(\mu ^{2}-\Delta ^{2})^{2}}{(\hbar v)^{6}})]  \label{Berry}
\end{equation}%
In Fig.~\ref{fig3} we show numerical results for $\Gamma _{+}(C)$ as a
function of warping $\lambda $ for various values of chemical potential $\mu
$ namely $\mu =0.4eV$ (black), $\mu =0.8eV$ (red), $\mu =1.2eV$ (green), $%
\mu =1.6eV$ (blue). In all cases $\Delta =0.1eV$. The solid curve are exact
numerical results based on Eq.~(\ref{Berry1}) while the dashed curves are
the approximate result of Eq.~(\ref{Berry}). We see that both sets agree
perfectly at small $\lambda $ but begin to deviate significantly as $\lambda
$ increases. In the limit of large $\lambda $ the Berry phase goes towards
its $\Delta =0$ value of $\pi $.

When $\lambda =0$%
\begin{equation}
\Gamma _{+}(C)=\pi (sign(\Delta )-\frac{\Delta }{\mu })
\label{Berry0}
\end{equation}%
which is a known result. Another known result can also be verified. If we
set $\hbar =1$ for simplicity and retain the Schr\"{o}dinger term $E_{0}(%
\mathbf{k})$ of Eq.~(\ref{Hamiltonian}) we can solve for $k_{F}$
\begin{equation}
k_{F}^{2}=2\mu m+2m^{2}v^{2}-2m^{2}\sqrt{2\frac{\mu }{m}v^{2}+v^{4}+\frac{%
\Delta ^{2}}{m^{2}}}
\end{equation}
and so%
\begin{eqnarray}
&&\sqrt{v^{2}k_{F}^{2}+\Delta ^{2}}=\mu -k_{F}^{2}/2m  \notag \\
&=&-mv^{2}+mv^{2}\sqrt{1+\frac{2\mu }{mv^{2}}+\frac{\Delta ^{2}}{(mv^{2})^{2}%
}}
\end{eqnarray}
which gives the Berry phase%
\begin{eqnarray}
&&\Gamma _{+}(C)=\pi \lbrack 1-\frac{\Delta }{\sqrt{v^{2}k_{F}^{2}+\Delta
^{2}}}]  \notag \\
&=&\pi \lbrack 1+\frac{\Delta }{mv^{2}[1-\sqrt{1+\frac{2\mu }{mv^{2}}+\frac{%
\Delta ^{2}}{(mv^{2})^{2}}}]}]\label{Berrym}
\end{eqnarray}
This result was obtained by Wright and Mckenzie.\cite{Wright1,Fuchs,Wright2}

\section{Spin texture}

\begin{figure}[tp]
\begin{center}
\includegraphics[height=3.0in,width=3.0in]{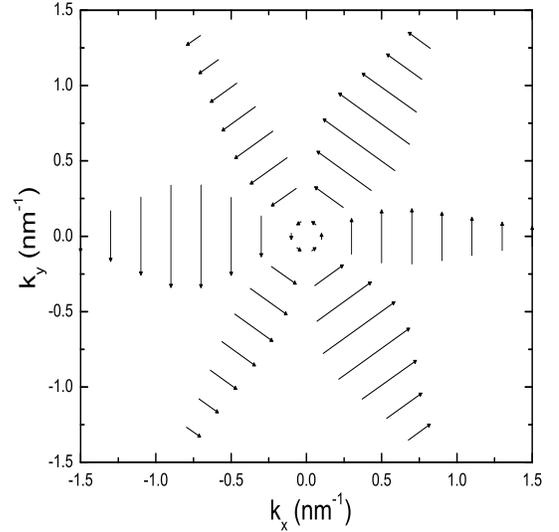}
\end{center}
\caption{Spin texture ($S_x$,$S_y$) in the $k_{x}$-$k_{y}$ plane of the
surface state Brillouin zone. The units of momentum are $nm^{-1}$. The gap $%
\Delta=0.1eV$ and the warping $\protect\lambda=0.2eV\cdot nm^{3}$.}
\label{fig4}
\end{figure}
\begin{figure}[tp]
\begin{center}
\includegraphics[height=4.4in,width=2.8in]{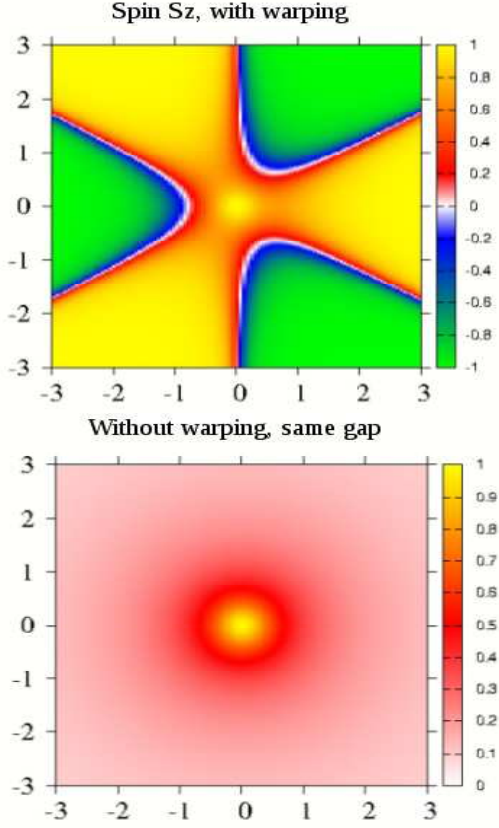}
\end{center}
\caption{(Color online) The z component of spin ($S_z$, in units of $\hbar/2$%
) perpendicular to surface as a function of $k_{x}$, $k_{y}$ in units of $%
nm^{-1}$ in the 2-D surface state Brillouin zone, with (top frame) and
without (bottom frame) warping. The gap $\Delta =0.1eV$ and in the top frame
the warping $\protect\lambda=0.2eV\cdot nm^{3}$. }
\label{fig5}
\end{figure}
The Pauli matrix in Eq.~(\ref{Hamiltonian}) refer to spin. From a knowledge
of the wave function given in Eq.~(\ref{wave}), we can compute the average
value of the electron spin components $S_{x}$, $S_{y}$ (in plane) and $S_{z}$
(out of plane). The results are%
\begin{eqnarray}
S_{x} &=&\frac{\hbar }{2}\langle u(\mathbf{k},\pm )|\sigma _{x}|u(\mathbf{k}%
,\pm )\rangle  \notag \\
&=&\pm \frac{\hbar }{2}\frac{-\hbar vk_{y}}{\sqrt{(\Delta +\lambda
(k_{x}^{3}-3k_{x}k_{y}^{2}))^{2}+\hbar ^{2}v^{2}k^{2}}}  \label{sx} \\
S_{y} &=&\frac{\hbar }{2}\langle u(\mathbf{k},\pm )|\sigma _{y}|u(\mathbf{k}%
,\pm )\rangle  \notag \\
&=&\pm \frac{\hbar }{2}\frac{\hbar vk_{x}}{\sqrt{(\Delta +\lambda
(k_{x}^{3}-3k_{x}k_{y}^{2}))^{2}+\hbar ^{2}v^{2}k^{2}}}  \label{sy} \\
S_{z} &=&\frac{\hbar }{2}\langle u(\mathbf{k},\pm )|\sigma _{z}|u(\mathbf{k}%
,\pm )\rangle  \notag \\
&=&\pm \frac{\hbar }{2}\frac{\Delta +\lambda (k_{x}^{3}-3k_{x}k_{y}^{2})}{%
\sqrt{(\Delta +\lambda (k_{x}^{3}-3k_{x}k_{y}^{2}))^{2}+\hbar ^{2}v^{2}k^{2}}%
}.  \label{sz}
\end{eqnarray}

\begin{figure}[tp]
\begin{center}
\includegraphics[height=3in,width=3.2in]{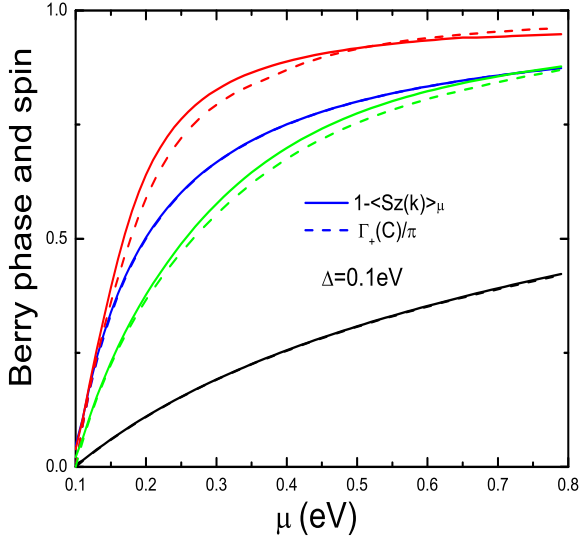}
\end{center}
\caption{(Color online) The Berry phase $\Gamma_{+}(C)$ of
Eq.~(\ref{Berryf}) normalized by $\pi$ (dashed lines) compared with
one minus the Fermi surface average z-component of spin
$\left\langle S_z \right\rangle_{\mu}$ defined in Eq.~(\ref{Szav})
(solid lines) as a function of chemical potential $\mu$. The blue
curves are for comparison and give results for the isotropic case
$\lambda=0$ (no warping). The Berry phase and spin agree perfectly
in this case. The gap is set at 0.1eV in all cases. The other curves
are for a warping of $\lambda=0.2eV(nm)^3$. The red curves have no
Schr\"{o}dinger quadratic piece ($E_0(k)=0$ in
Eq.~(\ref{Hamiltonian})), while the green are for $m=m_e$ (bare
electron mass) and the black for $m=0.1m_e$. The derivations between
dash and solid curves are always small and are reduced as $m$ is
decreased. } \label{fig6}
\end{figure}

We show our results in Fig.~\ref{fig4} for the $S_{x}$, $S_{y}$
component in the $k_{x}$, $k_{y}$ plane with momentum measured in
units of $nm^{-1}$. We see that the in-plane component of spin
remains locked perpendicular to its momentum
\cite{Hsieh1,Hsieh2,Xu1,Xu2} but its magnitude is no longer
independent of angle $\theta $. Basak et.al \cite{Basak} have gone
beyond the Hamiltonian (\ref{Hamiltonian}) to include a fifth order
spin-orbit coupling which could exist at the surface of a
rhombohedral crystal. This lifts the momentum spin locking found
here but goes beyond the present discussion. For $\lambda =0$ we
have
\begin{eqnarray}
S_{x} &=&\pm \frac{\hbar }{2}\frac{-\hbar vk_{y}}{\sqrt{\Delta ^{2}+\hbar
^{2}v^{2}k^{2}}}  \notag \\
S_{y} &=&\pm \frac{\hbar }{2}\frac{\hbar vk_{x}}{\sqrt{\Delta ^{2}+\hbar
^{2}v^{2}k^{2}}}.
\end{eqnarray}%
and hence
\begin{equation}
\sqrt{S_{x}^{2}+S_{y}^{2}}=\frac{\hbar }{2}\frac{\hbar vk}{\sqrt{\Delta
^{2}+\hbar ^{2}v^{2}k^{2}}}
\end{equation}%
which reduces to $\frac{\hbar }{2}$ independent of $k$ when there is no gap.
With a gap $\sqrt{S_{x}^{2}+S_{y}^{2}}=0$ at $k=0$ and saturates to value $%
\frac{\hbar }{2}$ at $k\rightarrow \infty $. These results imply that the z
component of spin is also modified by the presence of the gap and that there
are further modifications when $\lambda \neq 0$. Returning to the results of
Fig.~\ref{fig4} and Eq.~(\ref{sx}) and (\ref{sy}), we get for the magnitude
of the in plane component of spin with warping
\begin{equation}
\sqrt{S_{x}^{2}+S_{y}^{2}}=\frac{\hbar }{2}\frac{\hbar vk}{\sqrt{(\Delta
+\lambda (k_{x}^{3}-3k_{x}k_{y}^{2}))^{2}+\hbar ^{2}v^{2}k^{2}}}
\end{equation}%
For $k$ very large this no longer saturates to a value of $\frac{\hbar }{2}$
but tends to zero because of the $k^{3}$ dependence in the denominator. For $%
k\rightarrow 0$ however the in plane component of spin remains zero as we
found for the $\lambda =0$ case. Both these limiting behavior are clearly
seen in Fig.~\ref{fig4}. For a general value of $\mathbf{k}$ there is a
great deal of anisotropy in the magnitude of the in plane spin component. It
is maximum for $\theta =\pi /3$ or $\pi $ and minimum for $\theta =\pm 2\pi
/3$ and $0$. Turning next to the z-component of spin, note from Eq. (24)
that with $\Delta =0$ and $\lambda =0$, the z-component of spin $S_{z}=0$.
But $S_{z}$ is no longer zero when hexagonal warping is non-zero even for $%
\Delta =0$. The numerator in Eq. (24) is proportional to $\lambda
k^{3}cos(3\theta )$ and, as for the orbital magnetic moment we get zero only
along $\theta =\pm \pi /6$ and $\pi /2$. The z-component of spin is
otherwise finite and has regions where it is positive and other regions
where it is negative. There is no need to introduce magnetic dopants in the
system to see these effects. When a finite gap is opened the spin texture
for $S_{z}$ becomes even more complicated. In this case a color plot of the
value of the z-component of spin given in Eq.~(\ref{sz}) is presented in the
top frame of Fig.~\ref{fig5}. It has a maximum of $\frac{\hbar }{2}$ at $k=0$
where the in plane spin is zero and at larger $k$ values has three contours
of zeros, different from the contours of zero found in Fig.~\ref{fig1} for
the orbital magnetic moment. In the first case the zeros correspond to the
zeros of the equation $\Delta +\lambda k^{3}\cos (3\theta )=0$ and in the
second it is $\Delta -2\lambda k^{3}\cos (3\theta )=0$. In contrast for $%
\lambda =0$ the spin structure for the $S_{z}$ component of spin is
isotropic and given by
\begin{equation}
S_{z}=\pm \frac{\hbar }{2}\frac{\Delta }{\sqrt{\Delta ^{2}+\hbar
^{2}v^{2}k^{2}}}\label{sz0}
\end{equation}%
a color plot of the magnitude of $S_{z}$ as a function of $k_{x}$
and $k_{y}$ is found in the lower frame of Fig.~\ref{fig5} and is
for comparison with the top frame. Only the region near $k=0$ is
unaffected by the hexagonal warping. For the isotropic case
($\lambda =0$, no hexagonal warping) we can easily see that the
Berry phase $\Gamma _{+}(C)$ defined in Eq.~(\ref{Berry0}) for pure
Dirac and in Eq.~(\ref{Berrym}) when there is a mass term, are
reproduced when
the z-component of spin given by (\ref{sz0}) is averaged over the Fermi surface i.e.%
\begin{equation}
1-\Gamma _{+}(C)/\pi =\left\langle S_{z}\right\rangle _{\mu }\equiv \frac{%
\int d^{2}kS_{z}\delta (\mu -\varepsilon _{+}(\mathbf{k}))}{\int
d^{2}k\delta (\mu -\varepsilon _{+}(\mathbf{k}))} \label{Szav}
\end{equation}%
Even when warping is included we find numerically that this relationship
remains very nearly true as can be seen in Fig.~\ref{fig6} where we compare $%
1-\left\langle S_{z}\right\rangle _{\mu }$ (solid curve) with
$\Gamma _{+}(C)/\pi $ (dashed curve) as a function of chemical
potential $\mu $ for four cases. The blue curves are for reference
only and have no warping and no Schr\"{o}dinger. The red are for
m=infinity and $\lambda =0.2$. Dashed and solid curve deviate
slightly from each other but the correlation remains excellent. The
green and black curves have a mass $m=m_{e}$ (electron mass)
and $m=0.1m_{e}$ respectively. The deviation between Berry phase and $%
\left\langle S_{z}\right\rangle _{\mu }$ decreases with increasing $m$. A
measurement of $\Gamma _{+}(C)/\pi $ is equivalent to a measurement of $%
\left\langle S_{z}\right\rangle _{\mu }$ and vice versa.

\section{Circular polarization and dichroism}

To calculate the interband optical matrix element we need the velocity
components related to
\begin{subequations}
\begin{equation}
\frac{\partial H}{\partial k_{x}}=\left(
\begin{array}{cc}
\lambda (3k_{x}^{2}-3k_{y}^{2}) & -i\hbar v_{k} \\
i\hbar v_{k} & -\lambda (3k_{x}^{2}-3k_{y}^{2})%
\end{array}%
\right)
\end{equation}%
and
\begin{equation}
\frac{\partial H}{\partial k_{y}}=\left(
\begin{array}{cc}
-6\lambda k_{x}k_{y} & -\hbar v_{k} \\
-\hbar v_{k} & 6\lambda k_{x}k_{y}
\end{array}
\right).
\end{equation}
\end{subequations}
From this information we can work out the optical matrix element for
valence (v) to conduction band (c). By definition \cite{Xiao2,Yao}
\begin{equation}
\mathbf{P}^{cv}(\mathbf{k})=\left\langle u(\mathbf{k},+)|\nabla _{\mathbf{k}%
}H|u(\mathbf{k},-)\right\rangle.
\end{equation}
After considerable but straightforward algebra we find that the
square of the right and left polarization optical matrix elements
are
\begin{eqnarray}
|P_{+}^{cv}(\mathbf{k})|^{2} &=&\hbar ^{2}v^{2}\{[\frac{\Delta -2\lambda
(k_{x}^{3}-3k_{x}k_{y}^{2})}{\sqrt{\hbar ^{2}v^{2}k^{2}+(\Delta +\lambda
k_{x}^{3}-3\lambda k_{x}k_{y}^{2})^{2}}}+1]^{2}  \notag \\
&&+\frac{9\lambda ^{2}(k_{y}^{3}-3k_{x}^{2}k_{y})^{2}}{\hbar
^{2}v^{2}k^{2}+(\Delta +\lambda k_{x}^{3}-3\lambda k_{x}k_{y}^{2})^{2}}\}
\end{eqnarray}%
\begin{eqnarray}
|P_{-}^{cv}(\mathbf{k})|^{2} &=&\hbar ^{2}v^{2}\{[\frac{-\Delta +2\lambda
(k_{x}^{3}-3k_{x}k_{y}^{2})}{\sqrt{\hbar ^{2}v^{2}k^{2}+(\Delta +\lambda
k_{x}^{3}-3\lambda k_{x}k_{y}^{2})^{2}}}+1]^{2}  \notag \\
&&+\frac{9\lambda ^{2}(k_{y}^{3}-3k_{x}^{2}k_{y})^{2}}{\hbar
^{2}v^{2}k^{2}+(\Delta +\lambda k_{x}^{3}-3\lambda k_{x}k_{y}^{2})^{2}}\}
\end{eqnarray}%
The difference between $|P_{+}^{cv}(\mathbf{k})|^{2}$ and $|P_{-}^{cv}(%
\mathbf{k})|^{2}$ works out to be
\begin{eqnarray}
&&|P_{+}^{cv}(\mathbf{k})|^{2}-|P_{-}^{cv}(\mathbf{k})|^{2}  \notag \\
&=&4\hbar ^{2}v^{2}\frac{\Delta -2\lambda (k_{x}^{3}-3k_{x}k_{y}^{2})}{\sqrt{%
\hbar ^{2}v^{2}k^{2}+(\Delta +\lambda k_{x}^{3}-3\lambda k_{x}k_{y}^{2})^{2}}%
}\label{diff}
\end{eqnarray}%
which can be written in terms of the orbital magnetic moment Eq.~(\ref{mz})
as
\begin{equation}
|P_{+}^{cv}(\mathbf{k})|^{2}-|P_{-}^{cv}(\mathbf{k})|^{2}=-4\frac{\hbar }{e}%
m_{z}(\mathbf{k})[\varepsilon _{+}(\mathbf{k)-}\varepsilon _{-}(\mathbf{k)}]
\end{equation}%
We can also work out the sum of $|P_{+}^{cv}(\mathbf{k})|^{2}+|P_{-}^{cv}(%
\mathbf{k})|^{2}$, to get
\begin{eqnarray}
&&|P_{+}^{cv}(\mathbf{k})|^{2}+|P_{-}^{cv}(\mathbf{k})|^{2}  \notag \\
&=&2v^{2}\hbar ^{2}/[\hbar ^{2}v^{2}k^{2}+(\Delta +\lambda
k_{x}^{3}-3\lambda k_{x}k_{y}^{2})^{2}]\times   \notag \\
&&[\lambda
^{2}(5k_{x}^{6}+51k_{x}^{4}k_{y}^{2}-9k_{x}^{2}k_{y}^{4}+9k_{y}^{6})  \notag
\\
&&-2\Delta \lambda (k_{x}^{3}-3k_{x}k_{y}^{2})+2\Delta ^{2}+\hbar
^{2}v^{2}k^{2}]\label{sum}
\end{eqnarray}
which can be rewritten as
\begin{eqnarray}
| &&P_{+}^{cv}(\mathbf{k})|^{2}+|P_{-}^{cv}(\mathbf{k})|^{2}=[\varepsilon
_{+}(\mathbf{k)-}\varepsilon _{-}(\mathbf{k)]\times }  \notag \\
&&[\frac{\partial ^{2}[\varepsilon _{+}(\mathbf{k})-E_{0}(\mathbf{k})]}{%
\partial k_{x}^{2}}+\frac{\partial ^{2}[\varepsilon _{+}(\mathbf{k})-E_{0}(%
\mathbf{k})]}{\partial k_{y}^{2}}]
\end{eqnarray}

\begin{figure}[tp]
\begin{center}
\includegraphics[height=4.4in,width=2.4in]{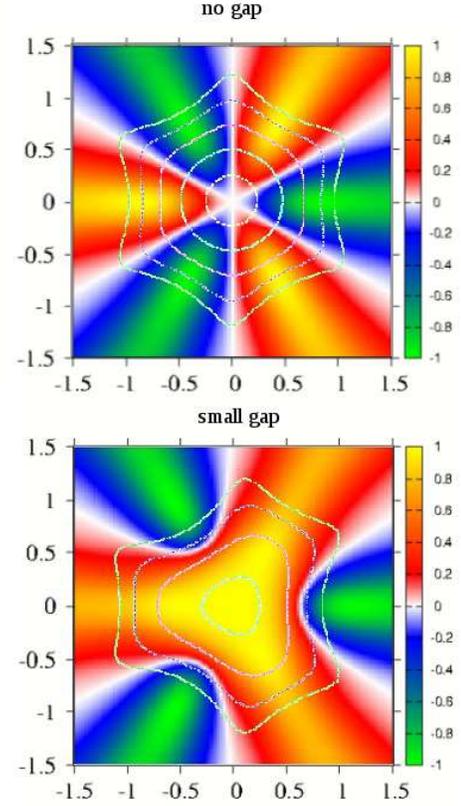}
\end{center}
\caption{(Color online) Color plots for the degree of circular polarization $%
\protect\eta(\mathbf{k})$ defined in Eq.~(\protect\ref{degree}) as a
function of $k_x$, $k_y$ in units of $nm^{-1}$ in the 2-D surface
state Brillouin zone. The contours (white lines) are added for
several values of the chemical potential (fermi surface). In the top
frame $\Delta=0$ and fermi contours go from circle to snowflakes. In
the lower frame the opening of a gap $\Delta=0.1eV$ distorts the
fermi contours.} \label{fig7}
\end{figure}

\begin{figure}[tp]
\begin{center}
\includegraphics[height=3in,width=3.2in]{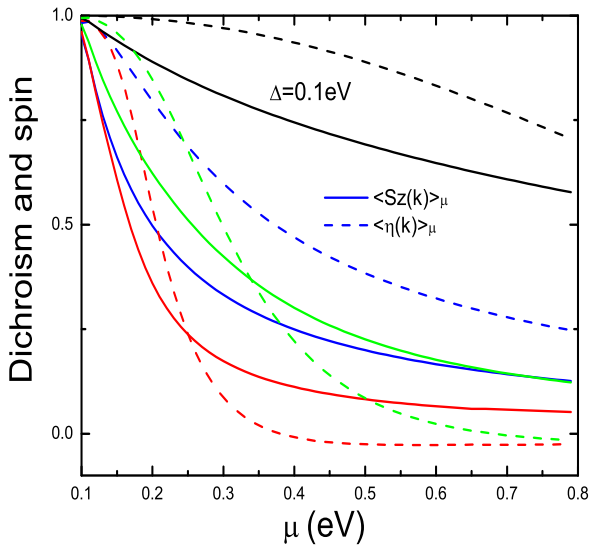}
\end{center}
\caption{(Color online) Comparison of dichroism $\left\langle\eta
(\mathbf{k})\right\rangle_{\mu}$ defined in Eq.~(\ref{eta}) (dashed
curves) and the Fermi surface averaged z-component of spin
$\left\langle S_z(\mathbf{k}) \right\rangle_{\mu}$ (solid curves)
defined in Eq.~(\ref{Szav}) as a function of chemical potential
$\mu$. The blue curves are for no Schr\"{o}dinger piece included in
the Hamiltonian (1) i.e. $E_0(k)=0$, and no hexagonal warping
($\lambda=0$) and are for comparison. There is some correlation
between spin-z and dichroism, both decrease monotonically with
increasing $\mu$ and at $\mu=0.8eV$ differ by a factor of 2. The red
curves have $\lambda=0.2eV(nm)^3$ as have all others and also have
$E_0(k)=0$. The green curves include a Schr\"{o}dinger piece with
$m=m_e$ and the black are for $m=0.1m_e$. In no case is the
correlation between spin and dichroism as good as that found in
Fig.~\ref{fig6} for spin and Berry phase. } \label{fig8}
\end{figure}

\begin{figure}[tp]
\begin{center}
\includegraphics[height=4.8in,width=3.0in]{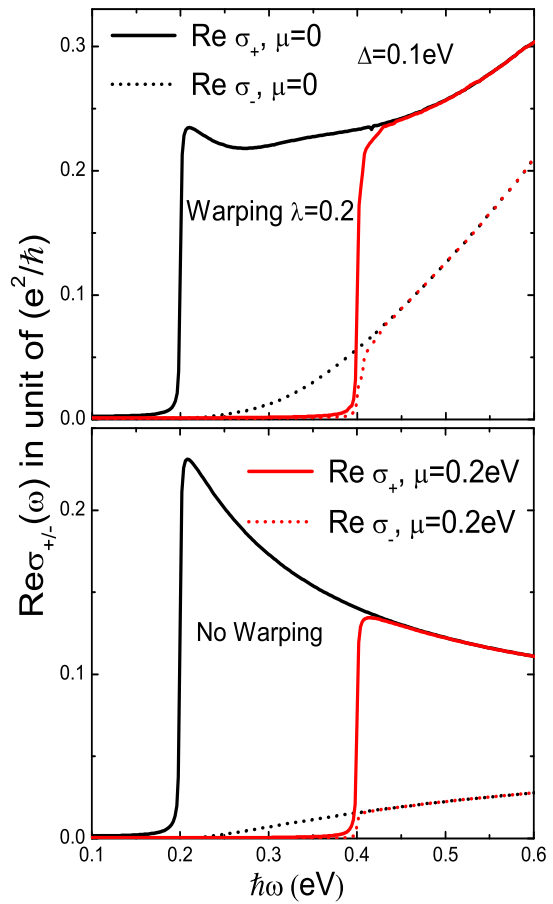}
\end{center}
\caption{(Color online) AC optical conductivity (in units of $e^2/\hbar$)
for circular polarized light as a function of photon energy $\protect\omega$
in eV. A gap of 0.1eV is included. The top frame includes a warping term of $%
0.2eV\cdot nm^{3}$ and the bottom frame has no warping and is for
comparison. Two value of chemical potential are shown as well as right $%
\protect\sigma_{+}$ (solid) and left hand $\protect\sigma_{-}$
(dotted) polarization.} \label{fig9}
\end{figure}

\begin{figure}[tp]
\begin{center}
\includegraphics[height=3in,width=3.2in]{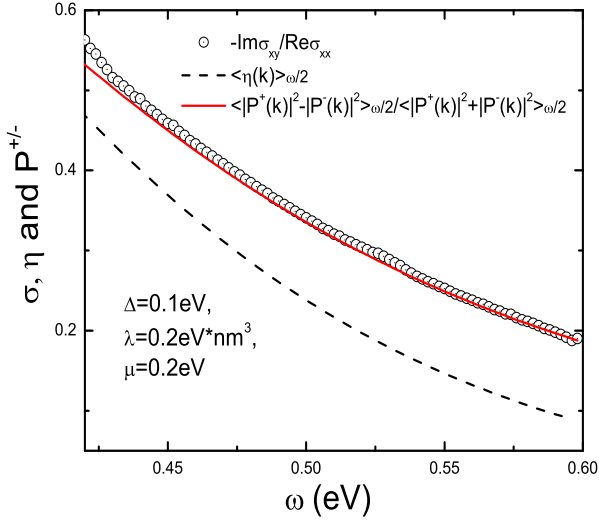}
\end{center}
\caption{(Color online) The ratio $-Im\sigma_{xy}/Re\sigma_{xx}$ of
Hall to longitudinal AC optical conductivity (open circles) as a
function of photon energy $\omega$ above the interband absorption
edge seen in Fig.~\ref{fig9}. The parameters are gap $\Delta=0.1eV$,
hexagonal warping $\lambda=0.2eV(nm)^3$ and chemical potential 0.2eV
as in Fig.~\ref{fig9}. Also shown as the dashed black curve is the
average of the dichroism factor $\eta (\mathbf{k})$ over a surface
of energy $\omega/2$. This quantity differs substantially from the
open circles. On the other hand, the red line agrees very well with
the open circles and is defined as the ratio of the average over a
surface of energy $\omega/2$ of $|P_{+}^{cv}(\mathbf{k})|^{2}-|P_{-}^{cv}(\mathbf{k%
})|^{2}$ and $|P_{+}^{cv}(\mathbf{k})|^{2}+|P_{-}^{cv}(\mathbf{k%
})|^{2}$ separately as defined in Eq.~(\ref{ome}). } \label{fig10}
\end{figure}

The degree of circular polarization is therefore given by
\cite{Xiao2,Yao,Ezawa}
\begin{eqnarray}
\eta (\mathbf{k}) &=&\frac{|P_{+}^{cv}(\mathbf{k})|^{2}-|P_{-}^{cv}(\mathbf{k%
})|^{2}}{|P_{+}^{cv}(\mathbf{k})|^{2}+|P_{-}^{cv}(\mathbf{k})|^{2}}  \notag
\\
&=&-\frac{-4\frac{\hbar }{e}m_{z}(\mathbf{k})}{[\frac{\partial
^{2}[\varepsilon _{+}(\mathbf{k})-E_{0}(\mathbf{k})]}{\partial k_{x}^{2}}+%
\frac{\partial ^{2}[\varepsilon _{+}(\mathbf{k})-E_{0}(\mathbf{k})]}{%
\partial k_{y}^{2}}]}  \label{degree}
\end{eqnarray}%
We note first that $\eta (\mathbf{k})$ is proportional to the orbital
magnetic moment and is further divided by the appropriate sum of the second
derivative of the energy $\varepsilon _{+}(\mathbf{k})-E_{0}(\mathbf{k})$
with respect to $k_{x}$ and $k_{y}$ respectively. While the energy $%
\varepsilon _{+}(\mathbf{k})$ depends on $E_{0}(\mathbf{k})$ the
quadratic Schr\"{o}dinger contribution to the Hamiltonian, the
difference $\varepsilon _{+}(\mathbf{k})-E_{0}(\mathbf{k})$ does not
and so the entire expression Eq.~(\ref{degree}) for $\eta
(\mathbf{k})$ is completely independent of this term. We show a
color plot for the magnitude of $\eta (\mathbf{k})$ as a function of
$k_{x}$, $k_{y}$ in the 2-D surface state Brillouin zone in Fig.~
\ref{fig7}. For a fix value of the absolute value of the momentum
the zeros in $\eta (\mathbf{k})$ (top frame) are at the same angles
as for the orbital magnetic moment of Fig.~\ref{fig1}. The top frame
of Fig.~\ref{fig7} is for
a case where $\Delta =0$ (no gap) while the bottom frame has a finite gap $%
\Delta =0.1eV$. In the case $\Delta =0$ averaging over angles at
fixed $k$ will give zero by symmetry because the numerator of
Eq.~(\ref{mz}) have the form $2\lambda k^{3}\cos (3\theta )$.
Similar results appear in Fig.~5 of a
paper by Liu et.al. \cite{Liu} Here we are mainly interested in the case when the gap $%
\Delta $ is non zero which is very different. When a gap is included the
numerator is instead $\Delta -2\lambda k^{3}\cos (3\theta )$, so that now we
get a non zero result proportional to the gap $\Delta $. Effectively the
hexagonal warping term has averaged to zero in the numerator leaving a
single term. This term however retains some knowledge of the warping as its
contribution remains in the denominator of Eq.~(\ref{degree}) and of Eq.~(%
\ref{mz}).

Interband optical absorption from a state $\mathbf{k}$ in the valence band
to a state $\mathbf{k}$ in the conduction band will involve $|P_{+}^{cv}(%
\mathbf{k})|^{2}$ and $|P_{-}^{cv}(\mathbf{k})|^{2}$ for right and
left circular polarized light respectively and $\eta (\mathbf{k})$
then measures the difference in absorption between these two cases.
For a given value of chemical potential $\mu >0$ there is an
absorption edge at $2\mu $ and the momentum involved is
$k_{F}(\theta )$ which we show in Fig.~\ref{fig7} as white contours
which start as circles for $\mu \rightarrow 0$ and distort to a more
and more cusped snowflake as $\mu $ is increased (top frame). When a
gap is included the Fermi surface distorts further as is shown in
the lower frame (white contours) and has reduced symmetry. At the
same time the color plot for $\eta (\mathbf{k})$ also shows reduced
symmetry as compared with the top frame. These anisotropies have
consequences for interband optical absorption.

It is useful to introduce an average over a constant energy surface
of the circular polarization $\eta (\mathbf{k})$ defined in
Eq.~(\ref{degree}) as this quantity is more closely connected with
the circular dichroism of the AC conductivity. We define
\begin{equation}
\langle \eta (\mathbf{k})\rangle _{\omega /2}\equiv \frac{\int d^{2}k\eta (%
\mathbf{k})\delta (\omega -2\varepsilon (\mathbf{k}))}{\int
d^{2}k\delta (\omega -2\varepsilon (\mathbf{k}))}\label{eta}
\end{equation}%
where $\varepsilon (\mathbf{k})=\sqrt{\hbar ^{2}v_{k}^{2}k^{2}+(\lambda
k^{3}\cos (3\theta )+\Delta )^{2}}$ which describes the energy involved in
interband transitions from valence to conduction band namely $2\varepsilon (%
\mathbf{k})\equiv \varepsilon _{+}(\mathbf{k})-\varepsilon
_{-}(\mathbf{k})$ where $\varepsilon _{\pm }(\mathbf{k})$ is defined
in Eq.~(\ref{disp}). Related
quantities are the separate averages of sum and difference $|P^{+}(\mathbf{k}%
)|^{2}$ and $|P^{-}(\mathbf{k})|^{2}$ defined in Eq.~(\ref{diff})
and (\ref{sum}) respectively. These are
\begin{eqnarray}
&&\langle |P^{+}(\mathbf{k})|^{2}\pm |P^{-}(\mathbf{k})|^{2}\rangle
_{\omega /2} \notag \\
&\equiv&\frac{\int d^{2}k[|P^{+}(\mathbf{k})|^{2}\pm |P^{-}(\mathbf{k}%
)|^{2}]\delta (\omega -2\varepsilon (\mathbf{k}))}{\int d^{2}k\delta
(\omega -2\varepsilon (\mathbf{k}))}\label{ome}
\end{eqnarray}%
Numerical results for the averaged circular dichroism $\langle \eta (\mathbf{%
k})\rangle _{\omega /2\equiv \mu }$ defined in (\ref{eta}) are
presented in Fig.~\ref{fig8} as dashed lines as a function of energy
$\mu $. In all cases the gap $\Delta
=0.1eV$. For the blue curve no Schr\"{o}dinger term is included i.e. $%
E_{0}(k)=0$ in the Hamiltonian (1), and there is no hexagonal warping ($%
\lambda =0$) and provides the isotropic limit as a reference for the
rest of the curves. It shows a monotonic decrease of $\langle \eta (\mathbf{k}%
)\rangle _{\omega /2}$ as a function of $\mu $. When a hexagonal warping
term of $\lambda =0.2eV\cdot nm^{2}$ is included we obtain the red curve
which shows that the dichroism is greatly reduced over its isotropic value
and is very small for values of $\mu $ greater than about $0.3eV$. If
however, in addition to the warping we include a Schr\"{o}dinger term with $%
m=m_{e}$ (with $m_{e}$ the bare electron mass) we obtain the green dashed
curve which has moved closer to the isotropic results than is the case for
the red curve. If we make the Schr\"{o}dinger piece $E_{0}(k)$ in (1) even
larger by taking $m=0.1m_{e}$ we get the black dashed curve which shows that
now $\langle \eta (\mathbf{k})\rangle _{\omega /2\equiv \mu }$ remains much
closer to one and is still of order $0.8$ at $\mu =0.8eV$.

It is of interest to compare these results for $\langle \eta (\mathbf{k}%
)\rangle _{\omega /2}$ with our previous results for the Fermi
surface average of the z-component of spin $\left\langle
S_{z}\right\rangle _{\mu }$ defined in Eq.~(\ref{Szav}). These are
shown as the solid curves in Fig.~\ref{fig8} which are color coded
to match the dashed curves. In all cases we note some correlation
between dichroism defined by Eq.~(\ref{eta}) and the average of the
z-component of spin. Both decrease monotonically with increasing
$\mu $ but in all cases there are significant quantitative
differences and contrary to what was found before when spin and
Berry phase were compared the dichroism does not translate directly
into a quantitative measure of $\left\langle S_{z}\right\rangle
_{\mu }$. Here we are describing the dichroism associated with the
absorption of light. Similar issues arise and have been studied
extensively when spin polarized angular resolved photoemission is
considered. \cite{Henk} In that case a photon enters the sample and
a photo electron is detected. Usually one concentrates on the
electron spectral density as the outcome of such experiments. But
the optical matrix element between incident light and electron also
enters and can have an effect on the ejected electron.
\cite{Herdt,Nomura,Yang}

To calculate in AC optical absorption we need to consider the Kubo formula
for longitudinal $\sigma _{xx}(\omega )$ and transverse $\sigma _{xy}(\omega
)$ conductivity. In terms of the matrix Green's function $\widehat{G}(%
\mathbf{k,}\omega _{l})$ in Matsubara notation with $\omega _{n}=(2n+1)\pi T$
and $\omega _{l}=2l\pi T$ the Fermion and Boson Matsubara frequencies, $n$
and $l$ are integers, $T$ is the temperature and $Tr$ is a trace, the
longitudinal conductivity for gapped Dirac fermion with warping is given by
\begin{eqnarray}
&&\sigma _{xx}(\omega )=\frac{e^{2}}{i\omega }\frac{1}{4\pi ^{2}}%
\int_{0}^{k_{cut}}kdkd\theta \times   \notag \\
&&T\sum_{l}Tr\langle v_{x}\widehat{G}(\mathbf{k,}\omega _{l})v_{x}\widehat{G}%
(\mathbf{k,}\omega _{n}+\omega _{l})\rangle _{\omega _{n}\rightarrow \omega
+i\delta }
\end{eqnarray}%
which works out to be
\begin{eqnarray}
&&\sigma _{xx}(\omega )=-\frac{e^{2}}{i\omega }\frac{1}{4\pi ^{2}}%
\int_{0}^{k_{cut}}\frac{kdkd\theta H(\theta )}{v_{k}^{2}k^{2}+(\Delta
+\lambda k^{3}\cos (3\theta ))^{2}}  \notag \\
&&[\frac{f(\varepsilon _{-})-f(\varepsilon _{+})}{\omega -2\sqrt{%
v_{k}^{2}k^{2}+(\Delta +\lambda k^{3}\cos (3\theta ))^{2}}+i\delta }  \notag
\\
&&-\frac{f(\varepsilon _{-})-f(\varepsilon _{+})}{\omega +2\sqrt{%
v_{k}^{2}k^{2}+(\Delta +\lambda k^{3}\cos (3\theta ))^{2}}+i\delta }]
\end{eqnarray}%
where
\begin{eqnarray}
&&H(\theta )=9\lambda ^{2}v_{k}^{2}k^{6}\cos ^{2}(2\theta
)+v_{k}^{4}k^{2}\sin ^{2}\theta   \notag \\
&&+v_{k}^{2}(\Delta +\lambda k^{3}\cos (3\theta ))^{2}  \notag \\
&&-2\times 3v_{k}^{2}\lambda k^{3}\cos (2\theta )\cos \theta \lbrack \Delta
+\lambda k^{3}\cos (3\theta )]
\end{eqnarray}%
From the matrix elements we have%
\begin{eqnarray}
&&\sigma _{xx}(\omega )=-\frac{e^{2}}{i\omega }\frac{1}{4\pi ^{2}}%
\int_{0}^{k_{cut}}kdkd\theta |P_{x}^{cv}(\mathbf{k})|^{2}  \notag \\
&&[\frac{f(\varepsilon _{-})-f(\varepsilon _{+})}{\omega -2\sqrt{%
v_{k}^{2}k^{2}+(\Delta +\lambda k^{3}\cos (3\theta ))^{2}}+i\delta }  \notag
\\
&&-\frac{f(\varepsilon _{-})-f(\varepsilon _{+})}{\omega +2\sqrt{%
v_{k}^{2}k^{2}+(\Delta +\lambda k^{3}\cos (3\theta ))^{2}}+i\delta }]
\end{eqnarray}%
The transverse Hall conductivity is given by
\begin{eqnarray}
&&\sigma _{xy}(\omega )=\frac{e^{2}}{i\omega }\frac{1}{4\pi ^{2}}%
\int_{0}^{k_{cut}}kdkd\theta \times   \notag \\
&&T\sum_{l}Tr\langle v_{x}\widehat{G}(\mathbf{k,}\omega _{l})v_{y}\widehat{G}%
(\mathbf{k,}\omega _{n}+\omega _{l})\rangle _{\omega _{n}\rightarrow \omega
+i\delta }
\end{eqnarray}%
which can be reduced to \cite{Xiao3}
\begin{eqnarray}
&&\sigma _{xy}(\omega )=\frac{e^{2}}{\omega }\frac{1}{4\pi ^{2}}%
\int_{0}^{k_{cut}}\frac{kdkd\theta v_{k}^{2}[\Delta -2\lambda k^{3}\cos
(3\theta )]}{\sqrt{v_{k}^{2}k^{2}+(\Delta +\lambda k^{3}\cos (3\theta ))^{2}}%
}  \notag \\
&&[\frac{f(\varepsilon _{-})-f(\varepsilon _{+})}{\omega -\varepsilon
_{+}+\varepsilon _{-}+i\delta }+\frac{f(\varepsilon _{-})-f(\varepsilon _{+})%
}{\omega -\varepsilon _{-}+\varepsilon _{+}+i\delta }]\}  \label{Hall}
\end{eqnarray}%
Results for the real part (absorptive part) of the circular polarized
conductivity [$\sigma _{\pm }(\omega )\equiv \sigma _{xx}(\omega )\pm
i\sigma _{xy}(\omega )$] $Re\sigma _{\pm }(\omega )$ are shown in Fig.~\ref%
{fig9}. Both frames include a gap $\Delta =0.1eV$ and the warping parameter $%
\lambda $ was set at $0.2eV\cdot nm^{3}$. The vertical axis is for $Re\sigma
_{\pm }(\omega )$ in units of $\frac{e^{2}}{\hbar }$ and the horizontal axis
is photon energy in units of ($eV$). The solid lines are for $Re\sigma
_{+}(\omega )$ while the dotted lines are for $Re\sigma _{-}(\omega )$,
black is for a chemical potential $\mu =0$ and red for $\mu =0.2eV$.
Comparing top and bottom frame we wish to emphasize two features. First,
hexagonal warping leads to an increasing absorption for $\omega $ greater
than its threshold value in the solid curves for right circular polarized
light $Re\sigma _{+}(\omega )$ (top frame) in contrast to the case without
warping (bottom frame) where a decreasing absorption is seen. Note that for $%
\mu =0$ the absorption edge is determined by the gap and falls at
$2\Delta =0.2eV$. For $\mu =0.2eV$ it is at $2\mu =0.4eV$. For left
circular polarized light it is the dotted curves which arise and
here again the absorption rises more sharply when hexagonal warping
is included. The curves with warping are concave up while those
without are concave down. A second feature that we wish to emphasize
is that the degree of dichroism is significantly affected by
hexagonal warping. For example in the red curves just above
threshold at $\omega =0.4eV$, the ratio of the absorption from right
to left polarized light is about $8$ when we neglect warping while
it decreases to $4$ with warping.

The ratio of the difference between $Re\sigma _{+}(\omega )$ and
$Re\sigma _{-}(\omega )$ normalized to its sum is related to the
Hall angle and is given by
\begin{equation}
\frac{Re\sigma _{+}(\omega )-Re\sigma _{-}(\omega )}{Re\sigma
_{+}(\omega )+Re\sigma _{-}(\omega )}=-\frac{Im\sigma _{xy}(\omega
)}{Re\sigma _{xx}(\omega )}\label{ratio}
\end{equation}
This ratio is shown as the open circles in Fig.~\ref{fig10} as a
function of photon energy $\omega $ for the parameters used in
Fig.~\ref{fig9} namely, a gap $\Delta =0.1eV$, a chemical potential
$\mu =0.2eV$ and a hexagonal warping parameter
$\lambda =0.2eV(nm)^{3}$. The dichroism is seen to decrease with increasing $%
\omega $ and has its maximum above the main interband absorption
edge of Fig.~\ref{fig9} at $\omega =2\mu =0.4eV$, where the ratio in
(\ref{ratio}) is about 0.6. These
results are closely related to the optical matrix elements defining $|P^{+}(%
\mathbf{k})|^{2}\ $and $|P^{-}(\mathbf{k})|^{2}$ of Eq.~(\ref{diff})
(difference) and (\ref{sum}) (their sum). Taking averages over a
constant energy surface of $\omega /2 $ as defined in
Eq.~(\ref{eta}) and (\ref{ome}), we get for $\langle \eta
(\mathbf{k})\rangle _{\omega /2}$ the black dashed curve which does
not agree well with the open circles. In particular note that these
differ by a factor of 2 at $\omega
=0.6eV$. On the other hand near perfect agreement is obtained when $|P^{-}(%
\mathbf{k})|^{2}-|P^{+}(\mathbf{k})|^{2}$ and $|P^{-}(\mathbf{k}%
)|^{2}+|P^{+}(\mathbf{k})|^{2}$ are separately averaged before their
ratio is taken which gives the solid red curve. It is clear from
this comparison that $\langle \eta (\mathbf{k})\rangle _{\omega /2}$
is not a good measure of the dichroism when there is hexagonal
warping. In this case numerator and denominator in the first
equality in (\ref{degree}) need to be separately averaged and then
their ratio taken.

We make one final point. The DC limit of the Hall conductivity $\sigma
_{xy}(\omega =0)$ follows from Eq.~(\ref{Hall}) and reduces to
\begin{equation}
\sigma _{xy}(\omega =0)=\frac{e^{2}}{\hbar }\int \frac{dk_{x}dk_{y}}{(2\pi
)^{2}}\Omega _{c}(\mathbf{k})[f(\varepsilon _{+})-f(\varepsilon _{-})\mathbf{%
]}
\end{equation}
with the Berry curvature given by Eq.~(\ref{curve}). If we place the
chemical potential at $\mu =0$ i.e. to fall in the gap the integral for the
Hall conductivity reduce to that given in Eq.~(\ref{phase}) for the Berry
phase where the integral goes to infinity. This gives
\begin{equation}
\sigma _{xy}(\omega =0)=\frac{e^{2}}{4\pi \hbar }
\end{equation}
which shows that the hexagonal warping term leaves the quantized value of
the Hall conductivity unchanged as we expect.

\section{Summary and Conclusions}

The presence of an hexagonal warping term in the Hamiltonian for the helical
Dirac electrons at the surface of a topological insulator changes the Fermi
surface to a snowflake shape at large values of chemical potential from a
circle at small values. Here we showed how this term changes the orbital
magnetic moment when a gap is also included. Without warping, the orbital magnetic moment $m_{z}(\mathbf{%
k})$ is isotropic in momentum space and directly proportional to the gap $%
\Delta $. With warping and no gap $m_{z}(\mathbf{k})$ is non-zero but its
value depends on the angle of $\mathbf{k}$ in the 2-D surface states
Brillouin zone with lines of zero along 3 directions, $\theta =\pm \pi /6$, $%
\pi /2$ thus changing sign six times. Its angular average at fix absolute
value of momentum $\mathbf{k}$ however vanishes. With a gap this cancelation
no longer holds and $m_{z}(\mathbf{k})$ recovers the behavior it has when
there is no warping as long as $\mathbf{k}$ is small. As $\mathbf{k}$
increases isotropy is lost and there are three contours along which $m_{z}(%
\mathbf{k})$ becomes zero and on crossing one of these contours
there is a change in sign. Hexagonal warping, as does the presence
of a gap, changes the Berry curvature and consequently the Berry
phase around a constant energy contour. Without warping, a known
result is that a gap reduces the Berry phase. Here we show that this
effect is reduced when there is warping and the phase returns to a
value close to $\pi $ as the strength of the warping $\lambda $ is
increased. The spin texture is also changed. The component of the
spin in the $\mathbf{k}$ plane remains locked perpendicular to its
momentum but now its magnitude is no longer $\hbar /2$ as it is when
the gap is zero but now varies with angle as well as with the
magnitude of $\mathbf{k}$.
With a gap but no warping $\lambda =0$, the pattern is isotropic in $\mathbf{%
k}$ space with magnitude of the spin in the ($x,y$) plane starting from zero
at $k=0$ and increasing to $\hbar /2$ at large $k$ such that $\hbar
^{2}v^{2}k^{2}\gg \Delta $. For $\lambda $ non zero the spin pattern is
anisotropic and depends on angle although it still starts at zero for $k=0$
and is isotropic in this limit. But as k increases, the pattern becomes
anisotropic and at large value of $k$ it tends towards zero rather than
saturate to one half as for the $\lambda =0$ case. The $z$-component of spin
follows a complimentary pattern as the total spin must be $\hbar /2$.
Without warping but with a gap, the value of $S_{z}$ starts at $%
\hbar /2$ at $k=0$, drops towards zero at large $k$ and is isotropic
independent of the angle of $\mathbf{k}$. For the finite $\lambda $
case however the magnitude of $S_{z}$ can become zero as the
magnitude of $\mathbf{k}$ increases in certain directions and can
change sign. In other directions there is no zero and no sign
changes. Thus warping introduces a rich spin texture not present
when it is neglected.

We find that the Fermi surface average of the z-component of spin
provides a quantitative measure of the Berry phase in all cases
considered here. This holds for warping as well as well as inclusion
of a subdominant non relativistic quadratic in momentum
Schr\"{o}dinger term in the Hamiltonian (1) in addition to the
dominant relativistic Dirac term. This piece introduces particle
hole asymmetry and provides corrections to both Berry phase and spin
texture.

The degree of circular dichroism is changed by the warping, as is
the square of the optical matrix element for interband absorption.
The degree of circular polarization is defined as the normalized
difference between right and left hand such interband matrix
elements. It follows closely the pattern in $\mathbf{k}$ space
established for the orbital magnetic moment. In fact it is
proportional to $m_{z}(\mathbf{k})$ with an additional denominator
which modulates its behavior slightly but provides no important
qualitative changes. These effects translate into changes in the
frequency dependent AC interband conductivity. Both longitudinal and
transverse (Hall) conductivity are altered and the degree of
dichroism is decreased. For example for a warping of $\lambda
=0.2eV\cdot nm^{3}$, a gap of $\Delta =0.1eV$ and a chemical
potential $\mu =0.2eV$, the ratio of the absorption from right to
left polarized light just above the absorption threshold decreases
from 8 to 4. In discussions of dichroism it is customary to relate
it to the ratio of the optical matrix elements defined in Eq.~(38)
and consider its dependence on momentum $\mathbf{k}$. Here however
$\eta(\mathbf{k})$ is an angular dependent quantity and so averaging
this ratio as in Eq.~(39) gives a different answer than averaging
numerator and denominator separately as in Eq.~(40) before taking
their ratio. We find that it is this second procedure that needs to
be used to get quantitative measure of the dichroism. Another
results is that while the dichroism is found to correlate
qualitatively with the constant energy surface average of the
z-component of spin, the correspondence between these two quantities
is not quantitative while it is between the Berry phase and
$\left\langle S_{z}(\mathbf{k})\right\rangle _{\omega/2}$

As has been stressed recently by Wright and Mckenzie and others
\cite{Wright1,Fuchs,Wright2} an essential element of topological
insulator surface states is the existence of particle-hole asymmetry
which results from the presence of a quadratic in momentum
Schr\"{o}dinger contribution to the Hamiltonian. This term does not
change the wave function although it modifies the energy. Thus the
Berry curvature, orbital magnetic moment and spin texture at zero
temperature are unaffected. On the other hand a quantity that is
averaged over a constant energy contour such as the Berry phase or
Fermi surface averaged z-component of spin is changed because the
contour itself depends on the Schr\"{o}dinger contribution to the
energy. Nonetheless the DC Hall conductivity remains quantized and
unchanged at a value of $\frac{e^{2}}{4\pi \hbar }$ when the
chemical potential $\mu =0$ falls in the gap. This is also true for
the warping which modifies the Berry curvature but leaves the value
of the DC Hall unchanged.

\begin{acknowledgments}
This work was supported by the Natural Sciences and Engineering
Research Council of Canada (NSERC), and the Canadian Institute for
Advanced Research (CIFAR).
\end{acknowledgments}

\end{document}